\documentstyle[psfig,emulateapj]{article}

\lefthead{Goldberg \& Natarajan}
\righthead{The Galaxy Octopole Moment as a Probe of Weak Lensing Shear
Fields}

\begin{document}

\title{The Galaxy Octopole Moment as a Probe of Weak Lensing Shear
Fields}

\author{David M. Goldberg and Priyamvada Natarajan}
\affil{Department of Astronomy, Yale University, New Haven, CT06511}

\begin{abstract}
In this paper, we introduce the octopole moment of the light
distribution in galaxies as a probe of the weak lensing shear field.
While traditional ellipticity estimates of the local shear derived
from the quadrupole moment are limited by the width of the intrinsic
ellipticity distribution of background galaxies, the dispersion in the
intrinsic octopole distribution is expected to be much smaller,
implying that the signal from this higher order moment is ultimately
limited by measurement noise, and not by intrinsic scatter.  We
present the computation of the octopole moment and show that current
observations are at the regime where the octopole estimates will soon
be able to contribute to the overall accuracy of the estimates of
local shear fields. Therefore, the prospects for this estimator from
future datasets like the Advanced Camera for Survey and the Next
Generation Space Telescope are very promising.
\end{abstract}
\keywords{}

\section{Introduction}

The analysis of weak shear fields of background galaxies lensed by
clusters has proved to be a rich field of inquiry (Blandford \&
Narayan, 1992; Kaiser, Squires \& Broadhurst, 1995, hereafter KSB;
Kaiser \& Squires, 1993; Mellier 1999; Bartelmann \& Schneider, 2001
and references therein for applications of weak lensing analysis to
observations). However, the underlying assumption of these analyses
has always been that all of the lensing information can be extracted
from combinations of the quadrupole moments of the light distribution,
which define an ellipticity and an orientation for the lensed arclets.
Under the assumption that the intrinsic orientation of the lensed
galaxy is random, the expectation value of the complex ellipticity can
be related to the reduced shear. Do combinations of the quadrupoles
contain all the useful information on the local shear field?
Refregier \& Bacon (2001) point out that any even moment of the light
distribution will be preferentially oriented along the shear even in
the weak lensing limit.

Moreover, it is generally taken for granted that strong lensing of
galaxies, whether producing multiple images or strong arcs, should be
analyzed in an fundamentally different way from weakly lensed
``arclets''.  Kneib et al. (1993), Bartelmann (1995), and others
describe the analysis of a strongly lensed arc which lies near a
critical curve, and how such images can be used to reconstruct the
lens potential.

It is not clear that the transition between arclet and arc is as
abrupt as the current mode of analysis would suggest.  Another way of
thinking about strong arcs is that lensing induces large octopole
moments in the light distribution.  This octopole has two components: a
skewness in the light distribution away from the lens, and a term
which expresses itself as an arc tangential to the surface of constant
shear.  Ideally, there is a transitional regime, one in which the
shear is sufficiently weak that the image would not be characterized
as an arc, but nevertheless, an octopole moment might be measurable.

The potential benefits of constructing this higher order estimator 
are significant.  Despite the anticipated noisiness of the octopole
signal, an optimal combination of this with the quadrupole signal
could significantly improve the overall signal to noise, since the
absolute value of the intrinsic octopole moments are expected to be
small compared to the quadrupole moments. Moreover, since the induced
octopole moments implicitly probe the radial variation of the shear
field, one may also get an independent local estimate of the radial
derivative of the reduced shear.

Forthcoming instruments, such as the Next Generation Space Telescope
(NGST), ongoing surveys, including the Deep Lens Survey (DLS), and
current and future cluster observations by the Hubble Space Telescope
(HST) using the Advanced Camera for Survey, will produce an avalanche
of data which need to be fully exploited.  The high-quality nature of
the data means that the limitation may due to large the number of galaxies
(crowded fields making shape measurements tricky) rather than the
quality of the imaging, and thus, the octopole
approach to data analysis may be quite fruitful.

Our approach in this paper is as follows.  The basic lensing formalism
of linear distortions and the notation used in this paper are outlined in
\S~\ref{sec:bg}. In \S~\ref{sec:oct} octopole moments are defined, in
the context of them arising from the second-order effects in a
circularly symmetric lens. In \S~\ref{sec:err}, the error analysis is
discussed, and it is shown that a combination of the quadrupole and
octopole can be used to estimate the reduced shear and its spatial
derivative.  Finally, in \S~\ref{sec:discuss}, we present a discussion
of future prospects, including application to observational data-sets.

\section{Background}
\label{sec:bg}

We begin by defining our convention and reminding the reader of the
standard weak lensing equations.  An excellent and exhaustive review
of this material can be found in Bartelmann \& Schneider (2001), from
which our conventions are borrowed.

A point mass, $M$, at a distance, $D_l$, will deflect a light beam coming
from a source at distance, $D_s$.  In particular, if the source is
observed at a 2-d angle $\vec{\theta}$ from the lens, then the beam
would be deflected by an angle, $\vec\alpha$:
\begin{equation}
\vec{\alpha}_{PS}=\frac{(D_l-D_s)D_l}{D_s}\frac{4GM}{c^2\theta}\hat\theta\
.
\label{eq:ptmass}
\end{equation}
More generally, if the lens is an extended object, with a surface mass
density, $ \Sigma ( \vec{\theta})$, then a dimensionless mass, the
convergence, may be defined as:
\begin{equation}
\kappa({\vec\theta})=\frac{\Sigma({\vec\theta})}{\Sigma_{cr}}\equiv
\frac{(D_l-D_s)D_l}{D_s}
\frac{4\pi G \Sigma({\vec\theta})}{c^2}\ .
\label{eq:kappa_def}
\end{equation}

The convergence may be thought of as a source term for a potential,
$\psi(\vec{\theta})$, and related via a Poisson-like equation:
\begin{equation}
\nabla^2 \psi(\vec{\theta})=2\kappa(\vec{\theta})\ ,
\label{eq:psi_def}
\end{equation}
where all gradients and divergences are calculated in the
two-dimensional $\theta$-space.

Extension of equation~(\ref{eq:ptmass}) to the continuous case, and
combination with equation~(\ref{eq:kappa_def}) yields a deflection
angle, 
\begin{equation}
\vec\alpha(\vec\theta)=\frac{1}{\pi}\int d^2\vec\theta' \kappa(\vec\theta')
\frac{\vec\theta-\vec\theta'}{|\vec\theta-\vec\theta'|^2}\ .
\label{eq:alpha_def}
\end{equation}

A beam observed at angle $\vec\theta$ must therefore have originated in the
source plane at
\begin{equation}
\vec\beta(\vec\theta)=\vec\theta-\vec\alpha(\vec\theta)\ ,
\end{equation}
under the thin lens approximation.

Suppose we observe a galaxy with its center of light at the position
$\vec\beta_0$ {\it in the source plane}. Throughout this paper, we
will define a local set of coordinates such that $\vec\beta_0$, and
correspondingly, $\vec\theta_0$ (the position of the source in the
lens plane) are at the origin. This assumed convention does not
change the final results, but merely makes the equations more compact.
Solving the lensing equation to linear order, one finds that both
$\vec{\beta}_0$ and $\vec{\theta}_0$ correspond to the center of mapping
(i.e. the first order lens mapping from source plane to image plane
preserves the position of the center of light).  

In linear analysis both $\vec\beta_0$ and $\vec\theta_0$ correspond to
the center of light. However, in the higher order analysis, discussed
below, if the position, $\vec\beta_0$, in the source plane is lensed
to the foreground, the corresponding lensed position will no longer
necessarily be the center of light.

Combining equations~(\ref{eq:kappa_def}) \& (\ref{eq:psi_def}), the
deflection angle, $\vec{\alpha}(\vec{\theta})$ can be written
explicitly as the gradient of the scalar potential,
$\vec\alpha(\vec\theta)\equiv\nabla_{\vec\theta} \psi$.  Since lensing
conserves surface brightness, a mapping from foreground to background
coordinates is sufficient to determine a background brightness map
from a foreground one (or vice-versa) provided a full knowledge of the
geometry of the system (cosmology plus the redshifts of the source and
lens) and mass distribution of the lens. Thus, we may expand around
the origin to determine a deprojection operator on a foreground light
distribution, which yields the amplification matrix,
\begin{equation}
{\bf A}({\vec{\theta}})\equiv
\frac{\partial{\vec{\beta}}}{\partial {\vec{\theta}}}=
\left( 
\delta_{ij}-\frac{\partial^2\psi({\vec{\theta}})}{\partial
\theta_i\partial \theta_j}\right)\equiv
\left(
\begin{array}{cc}
1-\kappa-\gamma_1 & -\gamma_2 \\
-\gamma_2 & 1-\kappa+\gamma_1
\end{array}
\right)\ .
\end{equation}
Rigorously speaking, this expression is the first term in a Taylor
series expansion of the distortion operator. The term $\gamma$ is a
complex shear term, representing the anisotropic part of the
distortion, with $\gamma=|\gamma|e^{2i\phi}$, and the real and
imaginary parts being denoted with the subscripts, ``1'' and ``2''
respectively, as per convention.  Using our locally defined
coordinate system, we have:
\begin{equation}
\beta_{i}={\bf A}_{ij}\theta_{j} 
\end{equation}
Likewise, assuming the absence of a caustic crossing
($1-\kappa-|\gamma| < 0$), this can be inverted uniquely to give a
projection function,
\begin{equation}
\theta_{i}={\bf A}_{ij}^{-1}\beta_{j}
\end{equation}

In this analysis, we focus on expanding this projection operator to
the next higher order to derive the octopole moment rather than
restricting ourselves to the quadrupole moment alone.  In general,
researchers have treated weak lensing fields in the manner described
by KSB or its variants (Bacon, Refregier \& Ellis, 2000; van Waerbeke
et al. 2001).  These techniques describe the mapping of source-plane
quadrupole light distributions to lens-plane distributions, and thus
use the observed ellipticity and an assumption of random orientation
to invert the shear field.  In this work, we aim to generalize these
transformations to the next higher order.  Our notation for the $n$-th
order moments of a galaxy is:
\begin{equation}
\langle x^n y^m \rangle \equiv \frac{
\int dx dy I(x,y) (x-\overline{x})^n (y-\overline{y})^m}{
\int dx dy I(x,y)}\ ,
\label{eq:momentdef}
\end{equation}
where an overline indicates the mean value, where for the quadrupole
$n+m=2$.  We may redefine the surface brightness, $I(x,y)$ to any
other function of flux without a loss of generality in the above
expression.

The quadrupole moments are generally translated into a complex
ellipticity via one of the relations:
\begin{equation}
\epsilon = \frac{\langle x^2\rangle -\langle y^2\rangle +2i\langle
xy\rangle}
{\langle x^2\rangle +\langle y^2\rangle+
2\sqrt{\langle x^2\rangle \langle y^2\rangle-\langle xy\rangle^2}}
\end{equation}
or 
\begin{equation}
\chi=\frac{\langle x^2\rangle -\langle y^2\rangle +2i\langle
xy\rangle} {\langle x^2\rangle +\langle y^2\rangle}
\end{equation}

The expectation value of the complex ellipticities around a circular
annulus gives an estimate of the combination of parameters:
\begin{equation}
E(\epsilon_\theta )\simeq g\equiv \frac{\gamma}{1-\kappa}\ ,
\end{equation}
\begin{equation}
E(\chi_\theta)\simeq 2g
\end{equation}
where $g$ is known as the reduced shear.  This expression implies a
degeneracy between values of the shear and convergence.  It should
also be noted that these relations hold only in the limit of small
variance in the intrinsic ellipticity distribution and for small
values of $g$.  Strictly speaking, these approximations are invalid 
in the strong lensing regime characterized by the dramatic arcs and 
multiple images.

\section{The Octopole Moments}
\label{sec:oct}

In this section, we address the issue of analysis of higher order
moments of the light distribution.  In particular, we show that, even
in the weak lensing limit in which one would strictly not expect to be
able to detect gravitational arcs, an arc-like signature can be found
from the octopole moments of the galaxy light distribution.

In order to simplify this discussion somewhat, we assume a radially
symmetric potential throughout, and for a fiducial galaxy we further
choose the coordinate system with a configuration such that it lies
along the positive x-axis. This is entirely equivalent to defining a
radial and tangential component.  However, for practical reasons, we
chose to do all calculations in Cartesian coordinates.

Under these assumptions, using only linear theory, the lensed
(subscript $\theta$) and unlensed (subscript $\beta$) moments can be
related in a very straightforward way since the Jacobian is diagonal.
Thus,
\begin{equation}
\langle x^n y^m\rangle_\theta=[A^{-1}_{11}]^n
[A^{-1}_{22}]^m
\langle x^n y^m\rangle_\beta
=\frac{1}{(1-\kappa-\gamma)^n}\frac{1}{(1-\kappa+\gamma)^m}
\langle x^n y^m\rangle_\beta\ .
\end{equation}
Application of this transform to the quadrupole terms in the weak
lensing limit, yields the complex ellipticity transformations above.
If galaxies are randomly oriented, as is assumed (however, see Crittenden
et al. 2001 for an estimate of the degree of expected intrinsic
alignments), then the expectation value of any of the intrinsic
octopole moments (and hence the lensed octopole moments in a
circularly symmetric potential) will necessarily vanish.

However, even in the absence of an intrinsic octopole moment, the
second order Taylor expansion of the lensing equation can give rise to
octopole moments.  Consider the limit when the source image has a
small but finite size, the lensed field may be approximated as:
\begin{equation}
\beta_i\simeq A_{ij}\theta_j+\frac{1}{2}A_{ij,k}{\theta_j}{\theta_k} \ .
\end{equation}
By inspection, the following symmetries: $A_{ik,j}=A_{ij,k}$ and
$A_{ji,k}=A_{ij,k}$, hold.  The corresponding terms (expressible
as local derivatives of the shear and convergence field) are:
\begin{eqnarray}
A_{11,1}&=&-\kappa_{,1}-\gamma_{1,1}=-2\gamma_{1,1}-\gamma_{2,2}
=-2\gamma_{1,1}-\frac{2\gamma}{r}=-2\gamma'-\frac{2\gamma}{r}\\ \nonumber
A_{12,1}&=&A_{21,1}=A_{11,2}=-\gamma_{2,1}=0 \\ \nonumber
A_{22,1}&=&A_{12,2}=A_{21,2}=-\kappa_{,1}+\gamma_{1,1}=-\gamma_{2,2}=-\frac{2\gamma}{r}\\ \nonumber
A_{22,2}&=&-\kappa_{,2}+\gamma_{1,2}=0 \ , 
\end{eqnarray}
where $\gamma'=\partial \gamma/\partial r$, the radial derivative of
the shear field.

Since the potential, and thus the surface density of the lens,
$\Sigma(\vec{\theta})$, is circularly symmetric, and the center of the
lens lies on the x-axis, $\kappa_{,2}=0$.  Likewise, since everywhere
on the x-axis, $\gamma_2=0$, $\gamma_{2,1}=0$.  In addition, we have
used the relation derived by Kaiser (1995):
\begin{equation}
\left( \begin{array}{c} 
\kappa_{,1} \\
\kappa_{,2}
\end{array}
\right)=
\left( \begin{array}{c} 
\gamma_{1,1}+\gamma_{2,2} \\
\gamma_{2,1}-\gamma_{1,2}
\end{array}
\right)\ .
\end{equation}
Finally, it can be shown that $\gamma_{2,2}=2\gamma/r$.

Summarizing and incorporating all the above symmetries and
simplifications, we have:
\begin{eqnarray}
x_\beta&\simeq&A_{11}x_\theta-\left[\gamma'+\frac{\gamma}{r}\right]x_\theta^2-\frac{\gamma}{r}y_\theta^2
\\ \nonumber
y_\beta&\simeq&A_{22}y_\theta-\frac{2\gamma}{r}x_\theta y_\theta
\end{eqnarray}

Thus, to second order in the position, the above equations may be
inverted:
\begin{eqnarray}
x_\theta&\simeq&A_{11}^{-1}x_\beta+A_{11}^{-3}\left[\gamma'+\frac{\gamma}{r}\right]x_\beta^2+A_{11}^{-1}A_{22}^{-2}\frac{\gamma}{r}y_\beta^2
\\ \nonumber y_\theta&\simeq&A_{22}^{-1}y_\beta+A_{11}^{-1}A_{22}^{-2}
\frac{2\gamma}{r}x_\beta y_\beta\ .
\end{eqnarray}

In order to compute the moments of the lensed field, we need to
transform the area element by determining the Jacobian consistently to
the same order,
\begin{eqnarray}
J &\equiv& \left| 
\begin{array}{cc}
\frac{\partial x_\theta}{\partial x_\beta} & \frac{\partial x_\theta}
{\partial y_\beta} \\
\frac{\partial x_\theta}{\partial x_\beta} & \frac{\partial
x_\theta}{\partial y_\beta}
\end{array}
\right| \\ \nonumber
&=&
 \left| 
\begin{array}{cc}
A_{11}^{-1}+2A_{11}^{-3}\left[\gamma'+\frac{\gamma}{r}\right]x_\beta & 2A_{11}^{-1}A_{22}^{-2}\frac{\gamma}{r}y_\beta \\
2A_{11}^{-1}A_{22}^{-2}\frac{\gamma}{r}y_\beta & A_{22}^{-1}+2A_{11}^{-1}A_{22}^{-2}\frac{\gamma}{r}x_\beta
\end{array}
\right|\\ \nonumber
&\simeq&
\mu+\left(2A_{11}^{-3}A_{22}^{-1}\left[\gamma'+\frac{\gamma}{r}\right]+
2A_{11}^{-2}A_{22}^{-2}\frac{\gamma}{r}\right)x_\beta\ , 
\end{eqnarray}
to first order in $x_\beta$.  This yields an area element in the
lensed field such that $dx_\theta dy_\theta=J dx_\beta dy_\beta$.

We may now compute the expected transformation of any given moment.  The
total flux, $f$ from the galaxy, for example is:
\begin{equation}
f_\theta=\int I(x,y) J dx_\beta dy_\beta=\mu f_\beta
\end{equation}
which is precisely the result in linear theory.
We may also compute the shift in the center of light compared to the
linear theory prediction.  Symmetry arguments, along with our
choice of coordinates clearly will produce no shift in the
y-coordinate direction.

However, 
\begin{eqnarray}
\langle x \rangle_\theta&=&
\frac{1}{\mu f_\beta} \int \left(
A_{11}^{-1}x_\beta+A_{11}^{-3}\left[\gamma'+\frac{\gamma}{r}\right]x_\beta^2+
A_{11}^{-1}A_{22}^{-2}\frac{\gamma}{r}y_\beta^2\right) 
J dx_\beta dy_\beta \\ \nonumber
&=&
\left[ 3A_{11}^{-3}\left(\gamma'+\frac{\gamma}{r}\right)+
2A_{11}^{-2}A_{22}^{-1}\frac{\gamma}{r}\right]\langle x^2
\rangle_\beta
+A_{11}^{-1}A_{22}^{-2}\frac{\gamma}{r}\langle y^2
\rangle_\beta \\ \nonumber
&=& \left[ 3A_{11}^{-1}\left(\gamma'+\frac{\gamma}{r}\right)+
2A_{22}^{-1}\frac{\gamma}{r}\right]\langle x^2
\rangle_\theta
+A_{22}^{-1}\frac{\gamma}{r}\langle y^2
\rangle_\theta
\end{eqnarray}

Since this shift merely reflects a change in position between the
first and second order approximation of the lensing inversion, it is
not directly measurable.  On the other hand, we may compute the
expectation values of the observables, $\langle x^3\rangle_\theta$ and
$\langle xy^2\rangle_\theta$:
\begin{eqnarray}
\langle x^3\rangle_\theta & =
& A_{11}^{-3}\langle x^3\rangle_\beta+
\left[ 2A_{22}^{-1}\frac{\gamma}{r}+
5A_{11}^{-1}\left(\gamma'+\frac{\gamma}{r}\right)
\right] 
\langle x^4\rangle_\theta 
+3A_{11}^{-1}\frac{\gamma}{r}
\langle x^2y^2\rangle_\theta\\ \nonumber
&-&\left[ 9A_{11}^{-1}\left(\gamma'+
\frac{\gamma}{r}\right)+
6A_{22}^{-1}\frac{\gamma}{r}\right]
\langle x^2 \rangle_\theta^2
-3A_{11}^{-1}\frac{\gamma}{r}
\langle x^2\rangle_\theta \langle y^2 \rangle^2_\theta  \\ \nonumber
 & =&
A_{11}^{-3}\langle x^3\rangle_\beta+
\left[ 2\frac{g}{r(1+g)}+
5\left(\frac{g'-g/r-g^2/r}{1-g^2}\right)
\right] 
\langle x^4\rangle_\theta 
+3\frac{g}{r(1-g)}
\langle x^2y^2\rangle_\theta\\ \nonumber
&-&\left[ 9\left(
\frac{g'-g/r-g^2/r}{1-g^2}
\right)+
6\frac{g}{r(1+g)}\right]
\langle x^2 \rangle_\theta^2
-3\frac{g}{r(1-g)}
\langle x^2\rangle_\theta \langle y^2 \rangle^2_\theta  
\label{eq:octx3}
\end{eqnarray}
and
\begin{eqnarray}
\langle xy^2\rangle_\theta&=&
A_{11}^{-1}A_{22}^{-2}\langle xy^2\rangle_\beta+
\left[
3A_{11}^{-1}\left(\gamma'+\frac{\gamma}{r}\right)+
6A_{22}^{-1}\frac{\gamma}{r} \right]
\langle x^2y^2\rangle_\theta+
A_{11}^{-1}\frac{\gamma}{r}
\langle y^4\rangle_\theta \\ \nonumber
&-&
\left[ 3A_{11}^{-1}\left(\gamma'+
\frac{\gamma}{r}\right)+
2A_{22}^{-1}\frac{\gamma}{r}\right]
\langle x^2 \rangle_\theta \langle y^2\rangle_\theta
-A_{11}^{-1}\frac{\gamma}{r}\langle y^2
\rangle^2_\theta \\ \nonumber
&=&
A_{11}^{-1}A_{22}^{-2}\langle xy^2\rangle_\beta+
\left[
3\left(\frac{g'-g/r-g^2/r}{1-g^2}\right)+
6\frac{g}{1+g} \right]
\langle x^2y^2\rangle_\theta+
\frac{g}{1+g}
\langle y^4\rangle_\theta \\ \nonumber
&-&
\left[ 3\left(\frac{g'-g/r-g^2/r}{1-g^2}\right)+
2\frac{g}{1+g}\right]
\langle x^2 \rangle_\theta \langle y^2\rangle_\theta
-\frac{g}{1-g}\langle y^2
\rangle^2_\theta
\label{eq:octxy2}
\end{eqnarray}
In both these expressions (eqns. 23 and 24), the terms proportional to
squares of the quadrupole moments arise from a second order shift in
the center of light.  By inspection, the expectation values of the
other two octopole moments, $\langle x^2 y\rangle_\theta$ \& $\langle
y^3 \rangle_\theta$, will vanish.

Note that with the exception of the term proportional to the intrinsic
octopole moments, both the lensed octopoles are a function of
observables: the second and fourth moments of the light, the reduced
shear, $g$, and its first derivative.  The octopole analysis, however,
retains the degeneracy found in quadrupole analysis.  However, since
the scatter in the intrinsic octopoles may be much smaller than the
scatter in the intrinsic ellipticity, and since it has a mean of zero,
extracting a profile of $g(r)$ from a finite number of noisy galaxies
may be done with potentially higher signal to noise by using the third
as well as the second moments.  We discuss this parameter extraction
in the next section.

\section{Error Analysis and Parameter Estimation}
\label{sec:err}

\subsection{Estimation of the Errors on the Moments}

The fact that lensing induces an octopole moment in an otherwise
elliptical galaxy is interesting, but by no means useful unless we are
actually able to measure this effect. We must measure both the octopole
moments and the hexadecipole moments with some accuracy in order to
extract the shear, and since both moments are much more likely to be
contaminated by sky noise than their quadrupole counterparts, it is
possible that measuring these moments may be quite difficult.  In
order to estimate the feasibility of such an investigation, we must
first consider the uncertainties in measuring the moments for a 
fiducial galaxy.

We make a number of simplifying assumptions in estimating the moment
uncertainties.  First, we assume that the only source of error arises
from the Poisson noise in the galaxy image and the background sky.  We
thus do not include the effects of a PSF, which are likely to be
considerably more important on the lower order quadrupole moments than
on the octo- and hexadecipoles. Similarly, we do not include
pixelation effects, which are also expected to be more important for
the quadrupoles.  However, we do assume that the background sky level
is constant (up to Poisson noise) over the entire image, and that the
galaxy image is unblended with any of its neighbors.  It is these
effects that will contribute significantly to the noise in the higher
order moments.

Given these constraints, we assume that a good estimate of the moments
of a galaxy can be achieved by integrating using a circular mask,
including a tophat and Gaussian.  Although, this is clearly not
optimal, as a circular mask will introduce biases in all moments, we
adopt this aperture for purposes of illustration in this work.  The
generality of the formalism developed here allows one to potentially
replace the surface brightness in equation~(\ref{eq:momentdef}), with
any other monotonic function of surface brightness.  However, optimal
determination of this function would rely on some knowledge of the
true radial light profile of the galaxy.

Clearly, the analysis can be successfully performed for an elliptical
mask, or even an arbitrarily shaped mask, in which instance the
relevant region above some given isophote could be determined
iteratively.  Discussion and application of optimal shape estimation
can be found in the literature on IMCAT (Kaiser), sExtractor (Bertin
\& Arnouts, 1996), FOCAS (Jarvis \& Tyson 1981).  However, these
pieces of software have focused on estimating only the ellipticity and
other shape parameters related to the quadrupole moments.

Specially promising among recent shape estimation methods is the
Shapelets technique (Refregier \& Bacon, 2001; Refregier, 2001) which
decomposes images into Hermite polynomial basis sets.  A potential
approach to computing octopole moments may be to determine the second
order shear operator directly in shapelet space.

However, it is beyond the scope of the analysis of this work to
investigate optimal techniques for measuring higher order moments.
This will be dealt with in future work.  For the present, we will
simply use the assumptions above to effectively characterize the
practical measurement uncertainties in typical images.

That said, one may approximate the uncertainty in $\langle x^n y^m
\rangle,\ \sigma_{n,m}$ as:
\begin{equation}
\sigma^2_{n,m}\simeq \frac{\int_0^{2\pi} d\theta \ \cos^{2n}(\theta) \
\sin^{2m}(\theta) \int^\infty_0 dr \ w^2(r,\theta) \ r^{1+2n+2m} \ 
[I(r,\theta)+ N]}
{\left( \int^{2\pi}_0 d\theta \int^\infty_0 dr \ w(r,\theta) r I(r,\theta) \right)^2}\ ,
\end{equation}
where $I(r,\theta)$ is the surface brightness in counts/area of the
galaxy, $N$ is the background sky brightness in counts/area, and
$w(r,\theta)$ is a mask.   For our analysis, we use two forms for this
mask, a tophat, with $w=1$ out to a specified outer radius, and a
circular Gaussian, as used in KSB.  

The form of the integral explicitly assumes that the light is very
nearly circularly distributed.  For an assumed radial profile, the
error may be approximated analytically or semi-analytically.  For our
analysis, we have used a de Vaucouleurs (1948) profile:
\begin{equation}
I(r,\theta)=I_e \exp \{-7.67[(R/R_e)^{1/4}-1]\}\ ,
\end{equation}
where $I_e$ is the characteristic brightness, and $R_e$ is the half
light radius.  We find similar error estimates for other assumed profiles.
Typically, we set the outer radius at 4 times the half-light radius.

Note that we ignore any covariances in the errors.  While inclusion of
these terms in our analysis would generally lower the implied errors,
their effect is expected to be small compared to other uncertainties
which we have explicitly overlooked.

In addition to measurement errors, both the ellipticity estimate of
the shear and the octopole measurement of the shear require an
estimate of the intrinsic scatter of the ellipticity or the
octopole.  The first we will label  $\sigma_{\chi}$, which is somewhat
different from the value normally given.  Since we assume a random
orientation, $\sigma_{\chi}$ represents the standard deviation in the
intrinsic distribution of the real part of the complex ellipticity, a
term which has an expected mean of zero.

The intrinsic variation in the octopole  moments in the octopoles are
labeled $\sigma_{\langle
x^3\rangle}$, and $\sigma_{\langle xy^2\rangle}$.  These terms
represent the standard deviation in the following dimensionless form:
\begin{equation}
\sigma^3_{\langle x^3\rangle}=\left< \left (\frac{\langle x^3
\rangle}{R_e^3} \right)^2 \right>^{1/2}\ ,
\end{equation}
with a similar form for the other octopole.  We again assume no
covariance.

To simplify the analysis, we assume that the half-light radius, $R_e$,
and the characteristic radius is known with perfect certainty, as is
the integrated flux.  Since these terms are dominant near the center
of the image, their errors will be considerably smaller than the
higher order moments.

\subsection{Uncertainty from Ellipticity Estimates}

Finally, we consider the uncertainty in the shear from the measurement
of the quadrupoles of a single galaxy.  In the limit of weak lensing,
the reduced shear may be approximated as:
\begin{equation}
g\simeq\frac{\chi_1-\chi_1^{(s)}}{2}=\frac{\langle x^2 \rangle -\langle
y^2 \rangle}{2r_0^2}-\frac{\chi_1^{(s)}}{2} \ .
\end{equation}
Since the orientation of the intrinsic ellipticity, $\chi_{\beta}$ is
random, $\sigma_{\chi_1}=\sigma_{\chi}/\sqrt{2}$.  
Ebbels et al. (2000) derive a probability distribution function
for the observed ellipticity distribution for galaxies of different
morphological types from the Medium Deep Survey.  Adapting this, we find a
reasonable value of $\sigma_\chi=0.30$.

Thus, the variance in the estimate of $g$ from the ellipticity is:
\begin{equation}
\sigma^2_{g-Q}\simeq\frac{\sigma^2_{2,0}+\sigma^2_{0,2}}{4r_0^4}+
\frac{\sigma_\chi^2}{4}
\end{equation}
In most instances, this uncertainty will be dominated by the spread in
the intrinsic galaxy ellipticities.

\subsection{Uncertainty from Octopole Estimates}

The details of parameter estimation from the octopole moments are
slightly more involved.  We may begin by simplifying the terms in
eqns.~(\ref{eq:octx3} and \ref{eq:octxy2}) which are functions of
$g$.  For small $g$, we may say:
\begin{equation}
\frac{g}{(1+\-g)}\simeq g
\ \ ; \ \ \frac{r g'-g-g^2}{1-g^2}\simeq r g'-g
\end{equation}

We then simplify eqns.~~(\ref{eq:octx3} and \ref{eq:octxy2}) to the
following expressions:
\begin{equation}
\langle x^3\rangle_\theta = A_{11}^{-3} \langle x^3\rangle_\beta +
u_{11} g + u_{12} r \ g'\ ,
\end{equation}
and
\begin{equation}
\langle xy^2\rangle_\theta=
A_{11}^{-1}A_{22}^{-2}\langle xy^2\rangle_\beta+
u_{21} g + u_{22}r\  g'\ ,
\end{equation}
where the terms, $u_{ij}$, can be derived by inspection, and their
uncertainties can be determined by the standard propagation of errors.
We reiterate here that this analysis assumes that the errors are
Gaussian and uncorrelated, which is certainly not the case.

Inverting these expressions yields:
\begin{equation}
g\simeq\frac{1}{u_{11}u_{22}-u_{12}u_{21}}
\left[ u_{22}(\langle xy^2\rangle_\theta-
A_{11}^{-1}A_{22}^{-2}\langle xy^2\rangle_\beta)-
u_{12}(\langle xy^2\rangle_\theta-
A_{11}^{-1}A_{22}^{-2}\langle xy^2\rangle_\beta)\right]
\end{equation}
and
\begin{equation}
r\ g'\simeq \frac{1}{u_{11}u_{22}-u_{12}u_{21}}
\left[ -u_{21}(\langle xy^2\rangle_\theta-
A_{11}^{-1}A_{22}^{-2}\langle xy^2\rangle_\beta)+
u_{11}(\langle xy^2\rangle_\theta-
A_{11}^{-1}A_{22}^{-2}\langle xy^2\rangle_\beta)\right]
\end{equation}
Since the expectation value of the intrinsic octopoles is zero,
estimating the parameters from these equations is straightforward.

Finally, from the form above, calculation of the variance and
covariance of $g$ and $r \ g'$ is a straightforward but tedious exercise
in propagation of errors.

\subsection{Results}

It now remains for us to compare errors estimated from the quadrupole
moments alone to those from the octopole moments.  For these
estimates, we assume the galaxy to intrinsically follow a de
Vaucouleurs profile.  Note that this assumption is necessary only in
computing the estimate of the error, and is not required for 
parameter estimation.  

The fiducial galaxy has a magnitude of 22 at $z=0.5$, and has a
half-light radius in the source plane of 2kpc.  Error estimates are
made for observations taken using the HST Wide Field camera using a
16.8 ks exposure as was used for observations of AC114 (Natarajan et
al. 1998).  Signal and sky counts were estimated using the HST WFPC2
exposure time calculator \footnote{\tt
http://www.stsci.edu/instruments/wfpc2/Wfpc2\_etc/wfpc2-etc.html}.
The lens and source are separated by 100 pixels in the image, and we
take a fiducial shear of $2\%$ at the center of the source. The lens
is assumed to be isothermal, with $2\%$ mean shear at that radius. The
precise amplitude of the shear is not that important in comparing the
ellipticity and octopole estimate error bars, since both techniques
are, in fact, first order in the shear.

The deviation in the intrinsic ellipticity, $\sigma_\chi$ is taken as
0.30, as discussed above while the scatter in both the intrinsic
octopoles is $0.1$.  The intrinsic octopole distribution is not well
studied, and understanding this may prove to be a fruitful
inquiry into galaxy shapes.  We select a very small scatter here for
the following reason.  If the intrinsic dispersion in the octopole
moments is sufficiently large that for most observational sets the
uncertainties in parameter estimation are dominated not by photon
Poisson noise, but by shot noise in the sample, then the octopole
moment at best improves the signal to noise compared to the quadrupole
by $\sqrt{2}$. The transition from Poisson to shot noise domination is 
illustrated in Fig. 2g.

\begin{figure}
\centerline{\psfig{figure=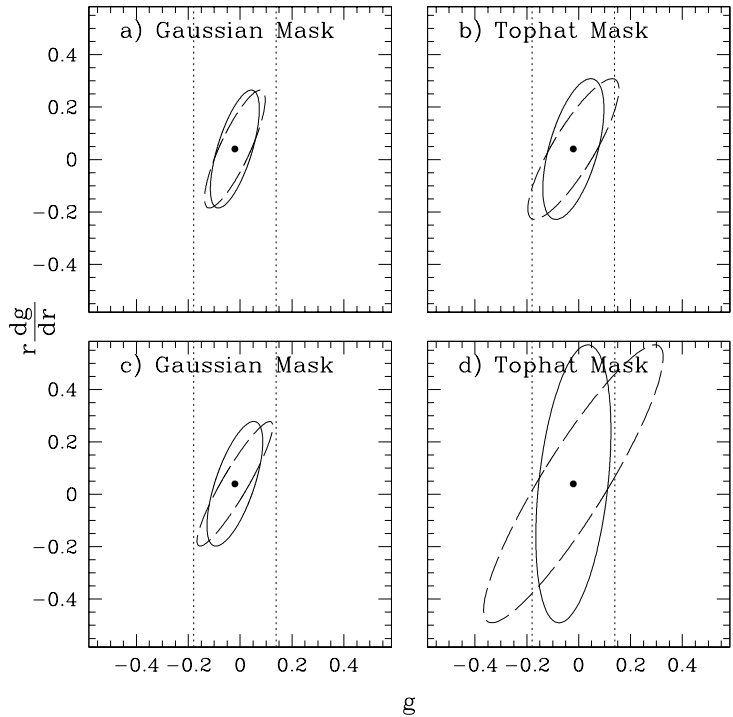,angle=0,height=4.0in}}
\caption{The calculated error ellipses using the method developed in
this paper. The dotted lines represent the $1\sigma$ errors on g from
the quadrupole technique alone. The dashed line represents the
$1-\sigma$ error ellipse on $g$ and $rg'$ from the octopole method.
The solid ellipse is the combined error estimate. In each panel, the
simulated galaxy has been assigned an apparent magnitude of 22 at $z=0.5$.
Calculations are done for the HST Wide Field Camera using the 814nm
filter.  Exposure time is set for 16.8ks, $\sigma_\chi=0.30$, and the
relative variances for both relevant octopole moments are set to 0.1.
The reduced shear is set to 2\%, and the source appears 100 pixels
from the lens.  The four panels use different masks to do the signal
to noise estimate.  Panels a) and c) use Gaussian masks, as described
in KSB, with characteristic radii set to the $R_e$ and $2R_e$,
respectively.  Panels b) and d) use tophat masks, with radii equal to
$2R_e$ and $4R_e$, respectively.  }
\label{fg:ellipse}
\end{figure}

The error ellipses from the quadrupole and octopole estimates of $g$
and $rg'$, along with their combined ellipse can be found in
Figure~\ref{fg:ellipse}.  Note that even for existing data sets,
inclusion of the octopole analysis has the potential to yield
significant new information about the shear field.

\begin{figure}
\centerline{\psfig{figure=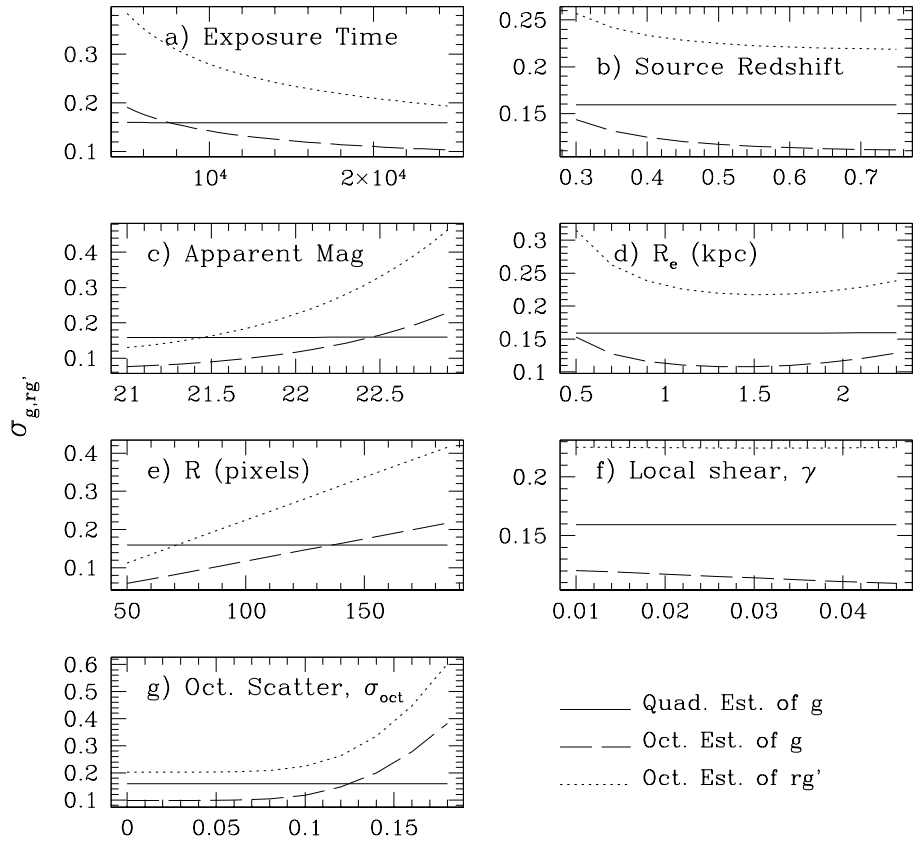,angle=0,height=5.0in}}
\caption{A sequence of slices through observational and physical
parameters, comparing the variation of uncertainty in parameter
estimates.  We vary a) exposure time, b) source redshift, c) source
apparent magnitude, d) source effective radius, e) source-lens
separation on the chip, f) local shear on the source, and g) intrinsic
scatter in the octopole.  Note that the uncertainty in $g$ arising from the
quadrupole estimate alone remains essentially unchanged over this set
of parameters.  This is due to the fact that  within this regime the 
dominant effect in the quadrupole signal is produced by shot noise.}
\label{fg:slice}
\end{figure}

The effect of the choice of physical and observational parameters on
the estimated errors are shown in Figure~\ref{fg:ellipse}. The error
ellipses for a slice through parameter space are shown in
Figure~\ref{fg:slice}. Analysis is performed using a Gaussian mask
with an effective radius equal to the half light radius.

Note that variations in the observational and physical parameters
within these ranges do not generally alter the error from the
quadrupole moment alone.  This is because in this regime, the
quadrupole error is dominated by the intrinsic variation in
ellipticities, rather than measurement errors.  If such is the case,
then we are necessarily limited by shot noise in the number of sources
rather than Poisson noise.  If the intrinsic scatter in the octopole
moment is sufficiently large, this may be the case for the octopole
estimator as well.  However, we reiterate that this statistic is not
well-studied.

For brighter sources or longer exposures, the quality of estimates for
the octopoles improves quickly, as do the estimates if the distance to
the lens decreases, even if this is not accompanied by a corresponding
increase in local shear.  

\section{Discussion and Future Prospects}
\label{sec:discuss}

In this paper, we have introduced a new and potentially powerful way
of analyzing weak lensing shear fields -- using the octopole moments
of the observed light distribution as a second-order estimator of
the shear. In addition to the shear, we also obtain an estimate
of its radial derivative. The radial derivative of the shear
field provides an important constraint on the mass profiles of
the dark matter halos that host galaxies. There are few other
methods that can probe that variation.          

While we have demonstrated that within the range of reasonable
physical and observational parameters the corresponding measurement
uncertainties for the octopole may be comparable to those found using
traditional ellipticity estimates of the shear, much remains to be
done, both theoretically and observationally.  The Gaussian mask used
in estimating the octopole and quadrupole moments is almost certainly
not optimal for this technique. In order to apply this to
observations, shape estimators must be used which can compute these
higher-order moments with maximum signal to noise. One avenue of
inquiry is to apply a second-order analysis to the Shapelet
(Refregier, 2001; Refregier \& Bacon, 2001) technique in order to get
a comparable signal.

In addition, we have created a somewhat simplistic model of the lens
as a circular system.  Schneider \& Bartelmann (1997), for example,
consider an extension of KSB with an elliptical lens. Such an effect
would introduce an additional degree of freedom in the present
analysis, since the shear and its radial derivative would no longer
necessarily align. Application of this technique with generality must
include these effects.

Finally, this technique needs to be applied to data.  A number of
excellent observational datasets exist with comparable observational
parameters to those used in the text (e.g. AC114, Natarajan et
al. 1998).  In addition, the Deep Lens Survey \footnote{\tt
http:\\dls.bell-labs.com} (DLS), and the forthcoming HST Advanced
Camera for Survey (ACS) will provide the sort of high-quality data
which will be ideal for this analysis. We hope to present an
application to currently available data-sets in a future paper.  It is
clear that there is significant information in higher-order moments of
the light distribution of galaxies, beyond just the ellipticity, and
here we have illustrated the uses of the octopole moment.

\acknowledgments

We would like to thank Ue-Li Pen and Tereasa Brainerd for helpful
discussions.

\end{document}